# Ordering of rare gas films on a decagonal Al-Ni-Co quasicrystal

R. D. DIEHL*[1], W. SETYAWAN[2], N. FERRALIS[1]†, R. A. TRASCA[1], M. W. COLE[1] AND S. CURTAROLO[2]

[1] Dept. Physics and Materials Research Institute, Penn State University,
University Park, PA 16802, USA
[2] Dept. Mechanical Engineering and Materials Science, Duke University,
Durham, NC, 27708, USA
† Current address: Dept. Chemical Eng., 201 Gilman Hall, University of California, Berkeley, CA 94720-1462

**Abstract**
This paper reviews recent progress in the study of rare gas films on quasicrystalline surfaces. The adsorption of Xe on the 10-fold surface of decagonal Al-Ni-Co was studied using low-energy electron diffraction (LEED). The results of these studies prompted the development of a theoretical model, which successfully reproduced the thermodynamic parameters found in the experiment. Grand canonical Monte Carlo (GCMC) simulations for Xe produced structures agreed with the experimental observations of the adsorption structures and provided a deeper insight into the nature of the ordering. A first-order commensurate-incommensurate transition that involves a transition from a quasicrystalline five-fold structure to a periodic hexagonal structure was discovered and characterized for the Xe monolayer. The five rotational domains of the hexagonal structure observed in the LEED study were shown in the GCMC study to be mediated by pentagonal defects that are entropic in nature, and not by substrate defects. The GCMC study found an absence of any such transition for Kr, Ar and Ne on the same surface. A detailed analysis of this transition led to the conclusion that the formation of the hexagonal layer depends on matching the gas and substrate characteristic lengths.

One motivation for studying the growth of films on quasicrystalline surfaces is the possible realization of new aperiodic structures, particularly those that contain a single element rather than the alloys normally associated with quasicrystals [1]. The tremendous variety of structures observed already in thin film growth on quasicrystal surfaces [2] indicates that these structures depend on a delicate balance of the adsorption interactions, which include the various bonds that form between the adatoms and substrate atoms, as well as inter-adatom interactions. Because of the complexity of the chemical bonding involved in metal films, we recently embarked on a series of experimental and theoretical studies of rare gas adsorption on quasicrystal surfaces in order to gain insight into the physical parameters that govern their growth [3-6]. The relative simplicity of rare gas interactions and the vast experience in modeling rare gas adsorption provides an ideal basis for understanding the effects of quasicrystallinity on adsorption.

Low-energy electron diffraction (LEED) experiments for Xe adsorbed on the 10-fold surface of d-Al-Ni-Co [3] found that Xe adsorbs layer-by-layer on the surface. The ordering of the first monolayer could not be established because LEED probes several layers [7], but non-quasicrystalline ordering was observed to occur at the onset of second layer adsorption. The bilayer structure was determined to consist of two layers of hexagonal Xe, with an average nearest-neighbor spacing of 0.44 nm. This hexagonal structure has a definite alignment with the substrate. Five different rotational domains are present, each having one of its six-fold axes aligned parallel to a different five-fold direction of the quasicrystal. At higher coverages, the film structure was found to be that of Xe(111) (still with five domains). The heat of adsorption of the monolayer Xe was measured to be 250 meV, which is similar to values obtained for Xe adsorption on a variety of metal surfaces [8].

This system was modeled using Lennard-Jones potentials for the Xe-Xe and Xe-substrate interactions [9]. The interaction of the Xe with the Al-Ni-Co substrate was obtained by summing pair potentials for a Xe atom and all of the substrate atoms in an 8-layer slab. The positions of the atoms in the 8-layer slab were taken from the results of a LEED analysis of the surface structure of Al-Ni-Co [10]. The Lennard-Jones parameters for the pairs were derived using conventional combining rules and experimental heats of adsorption [9]. The resulting Xe-substrate potential is shown in Figure 1. A similar procedure was followed to obtain the potentials for Kr, Ar and Ne [11]. A common feature of all of these potentials is that they are relatively deep and highly corrugated compared to rare gas potentials for other metal surfaces. Table 1 shows the range of adsorption energies for each of the rare gases, compared to values for Xe adsorption on a few other surfaces.

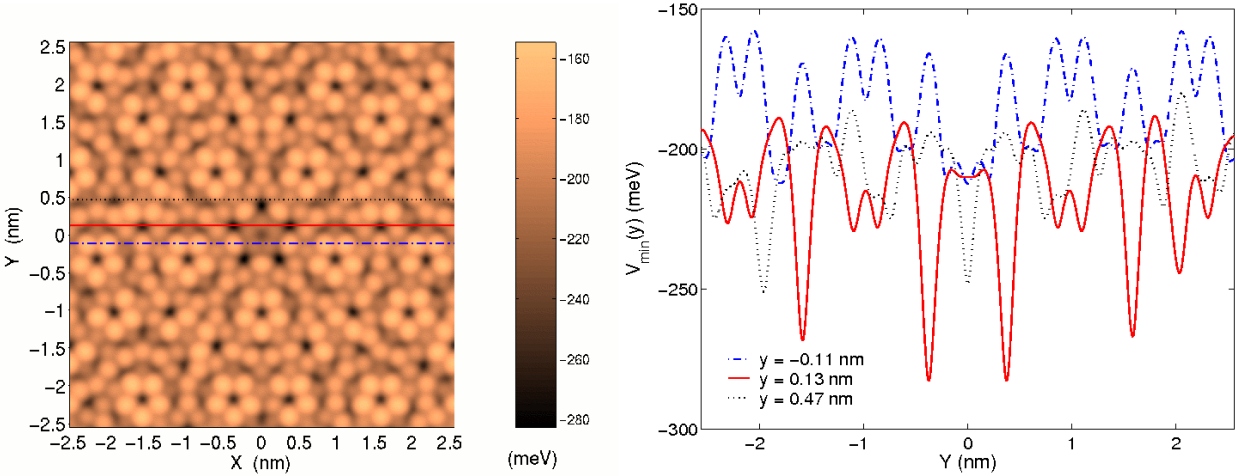

**Figure 1.** (color online). (a) Minimum potential energy for a Xe atom above an Al-Ni-Co surface, calculated using Lennard-Jones potentials. The size of the square region is 5.6 x 5.6 nm. (b) Line profiles (locations shown in (a)) for the potential energies of three parallel paths across the surface. The profiles at 0.13 nm and 0.47 nm indicate paths of relatively low average potential energy, while the profile at -0.11 nm indicates a path of relatively high average potential energy.

**Table 1.** Adsorption well depths and corrugations for selected rare gas adsorption systems. The $V_{min}$ Range column refers to the range of the potential energy minima found at different locations on the surface. $<V_{min}>$ and $\Delta V_{min}$ refer to the average well depth and the corrugation amplitude, respectively. The entries for single metal substrates are taken from a density functional theory (DFT) study of Xe adsorption [13]. The well depths for those are the experimental values reported in that paper, and the corrugation is the DFT-LDA calculated difference in energy between the top site and fcc hollow site.

| Substrate | Gas | $V_{min}$ Range (meV) | $<V_{min}>$ (meV) | $\Delta V_{min}$ (meV) |
|---|---|---|---|---|
| Al-Ni-Co | Ne | -71 to -33 | 47 | 38 |
| Al-Ni-Co | Ar | -181 to -85 | 113 | 96 |
| Al-Ni-Co | Kr | -225 to -111 | 146 | 114 |
| Al-Ni-Co | Xe | -283 to -155 | 195 | 128 |
| Cu(111) | Xe | ---- | 190 | 9 |
| Pt(111) | Xe | ---- | 310 | 49 |
| Pd(111) | Xe | ---- | 360 | 51 |

The submonolayer adsorption of Xe and Ar was first studied using the virial expansion equation of state, applicable in the limit of low coverage (Henry's Law regime) and at a somewhat higher coverage where the interadsorbate interactions become appreciable [9]. For the case of Xe, the temperature (T) and pressure (P) conditions found for the onset of adsorption agree well with the experimental values obtained from equilibrium isobars.

The large corrugation of the potential was found to have a significant effect on the adsorption behavior. An examination of the second virial coefficients for Xe and Ar indicates that the interaction between the adsorbed atoms is significantly more attractive than on a flat surface having the same average holding potential. This is because the corrugation brings the adatoms closer together than on a flat surface. This increased attraction also contributes to a larger heat of adsorption. The calculated heat of adsorption for Xe at a coverage of 0.25 and T= 70 K is 305 meV, close to the experimental value under the same conditions (300 meV) [9]. Therefore, we have a high degree of confidence that the calculated potentials are accurately describing the experimental system.

In order to gain insight into the ordering of the rare gas films, grand canonical Monte Carlo (GCMC) simulations were performed using the calculated potentials. One question that arose from the experiments on Xe adsorption concerns the order of the monolayer: is it quasiperiodic or disordered? A second question concerned the relative ease with which the film adopted a hexagonal structure at the onset of the adsorption of the second layer, especially considering the highly corrugated potential, which does not have hexagonal symmetry. As described below, both questions were addressed in the simulations.

In the GCMC simulations, the adsorption behavior of the rare gases on the Al-Ni-Co was first characterized using adsorption isotherms, as shown in Figure 2. Although vertical risers (steps) are evident in the isotherms, indicating layer-by-layer growth of the rare gases, the shapes of the isotherms have characteristics that differ from those obtained from typical periodic substrates[12]. In particular, the plateaus between the steps corresponding to first-layer and second-layer adsorption are not as horizontal as subsequent plateaus, and furthermore, their slopes increase with the declining size of the rare gas atoms. Both observations are consistent with the large corrugation of the substrate potential. The flatness of the subsequent plateaus indicates that gas monolayer produce flatter (less variable potential energy) surfaces for the adsorption of subsequent layers. Furthermore, the corrugation of the substrate potential has a larger effect on the smaller gas atoms, as might be expected, since they penetrate farther into the potential energy "holes", that are as large as 0.4 nm wide.

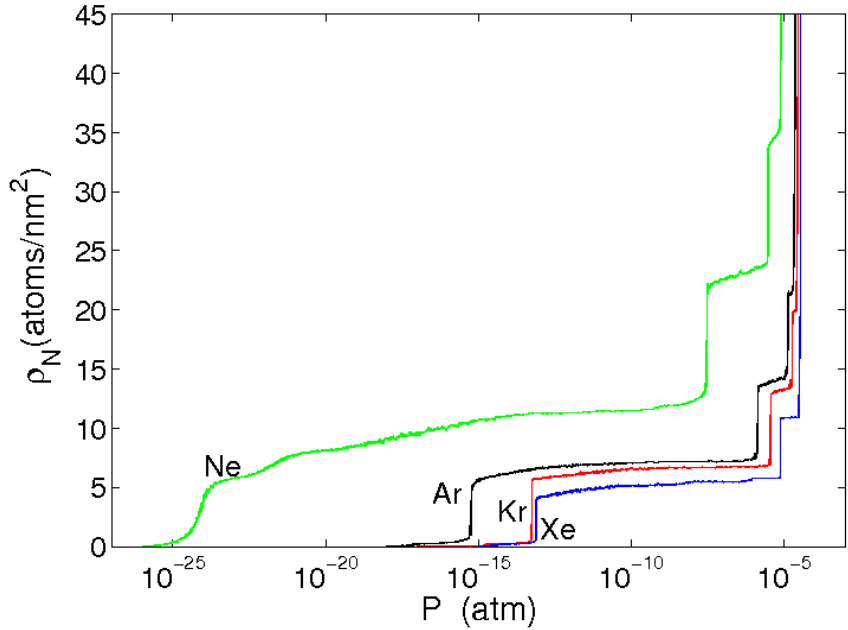

**Figure 2.** (color online). Calculated isotherms for Ne, Ar, Kr and Xe on Al-Ni-Co, showing adsorbed density vs. P. The T's of the isotherms are chosen so that T = 0.35e, where e is the Lennard-Jones gas phase interaction parameter. The actual T's are 11.8 K, 41.7 K, 59.6 K and 77 K, respectively. Note the large density variation in the first (low P) plateau relative to later (high P) plateaus, and the large density variation for the smaller gases relative to the larger gases.

We can see effect of the corrugation on the adsorption distance of the gases if we compare the average perpendicular height of the gas atoms, $d_{gs}$, on the quasicrystal surface to the same parameter on a flat surface having the same average holding potential, as shown in Table 2. For all gases, $d_{gs}$ is smaller than on the flat surface. The larger effect on the smaller gases can be seen by normalizing difference between these values to the size of the gas atom, as shown in the last column of Table 2. The difference relative to the gas size is larger for the smaller gases, indicating a larger relative effect of the corrugation on the smaller gases. Table 2 also gives similar values obtained from the obtained from LEED studies for Xe on metal surfaces, which are far less corrugated than the QC (see Table 1). The $d_{gs}$ values for these are considerably larger, except in the case of Pd(111), which has a significant chemical component within the gas-substrate bond [13].

**Table 2.** Average adsorption height $d_{gs}$ for the first layer of adsorbed rare gases. For the Al-Ni-Co this was determined from the GCMC simulations. For the single metal substrates, it was determined from LEED studies for the commensurate $(\sqrt{3}\times\sqrt{3})R30°$ structures [15]. d(g-flat) is the calculated adsorption height for the gases on flat substrates having the same average holding potential as the corrugated surface. $\Delta$ is the difference between the d(g-flat) and $d_{gs}$. $\sigma_{gg}$ is the Lennard-Jones parameter for the gas.

| substrate | gas | d(g-flat) | $d_{gs}$ | $\Delta$ | $\sigma_{gg}$ | $\Delta/\sigma_{gg}$ |
|---|---|---|---|---|---|---|
| Al-Ni-Co | Ne | 2.6107 | 2.3578 | 0.0657 | 2.78 | 0.8481 |
| Al-Ni-Co | Ar | 2.9131 | 2.6017 | 0.071 | 3.4 | 0.7652 |
| Al-Ni-Co | Kr | 3.0098 | 2.7125 | 0.0631 | 3.6 | 0.7535 |
| Al-Ni-Co | Xe | 3.2510 | 3.0456 | 0.033 | 4.1 | 0.7428 |
| Cu(111) | Xe | | 3.60 | | 4.1 | |
| Pt(111) | Xe | | 3.4 | | 4.1 | |
| Pd(111) | Xe | | 3.07 | | 4.1 | |

As might be expected from systems having similar values for gas-gas interaction energy and the lateral variation in potential energy, the simulations indicate that Xe adsorption on Al-Ni-Co is rich with phenomena arising from competing interactions [4-6]. The first structure that forms upon adsorption of the monolayer is indeed ordered and quasicrystalline, having the same symmetry as the substrate. As the monolayer compresses (through further adsorption), it undergoes a first-order phase transition into a hexagonal monolayer. The exact conditions (T, P) under which this transition occurs vary somewhat, as discussed below, but it always occurs before the onset of the second layer. This finding is somewhat at variance with the experimental finding, for which the simplest interpretation is that there is a continuous conversion to six-fold order as the second layer forms, suggesting that the second layer is required for the conversion. However, we note that it is difficult to pinpoint the onset of the hexagonal ordering in the experiment. Nevertheless, the transition of Xe to a six-fold structure occurs in both the experiment and the simulation, whereas it is not observed in simulations for other gases, as described below. An interesting feature of the six-fold structure that was evident in the simulations is that the different rotational domains present on the surface are a consequence of pentagonal defects present in the overlayer, and not due to substrate defects [5, 6].

The phase transition from the 5-fold quasicrystalline structure to the 6-fold hexagonal structure was characterized by defining an order parameter $\rho_{5-6}$:

$$\rho_{5-6} \equiv \frac{N_5}{N_5 + N_6},$$

where $N_5$ and $N_6$ are the numbers of atoms having 2D coordination equal to 5 and 6, respectively [6, 11]. Figure 3 shows this order parameter for Xe as a function of the reduced chemical potential:

$$\mu^* \equiv \frac{\mu - \mu_1}{\mu_2 - \mu_1},$$

where $\mu_1$ and $\mu_2$ are the chemical potentials of the onset of the first and second layer formation, respectively. The sharp drop in the order parameter near $\mu^* = 0.8$ is indicative of a first-order phase transition. This structural phase transition can be thought of as a commensurate-incommensurate transition, since the structure at low µ is commensurate with the substrate, whereas the structure at high µ is incommensurate, and furthermore has a different symmetry than the substrate.

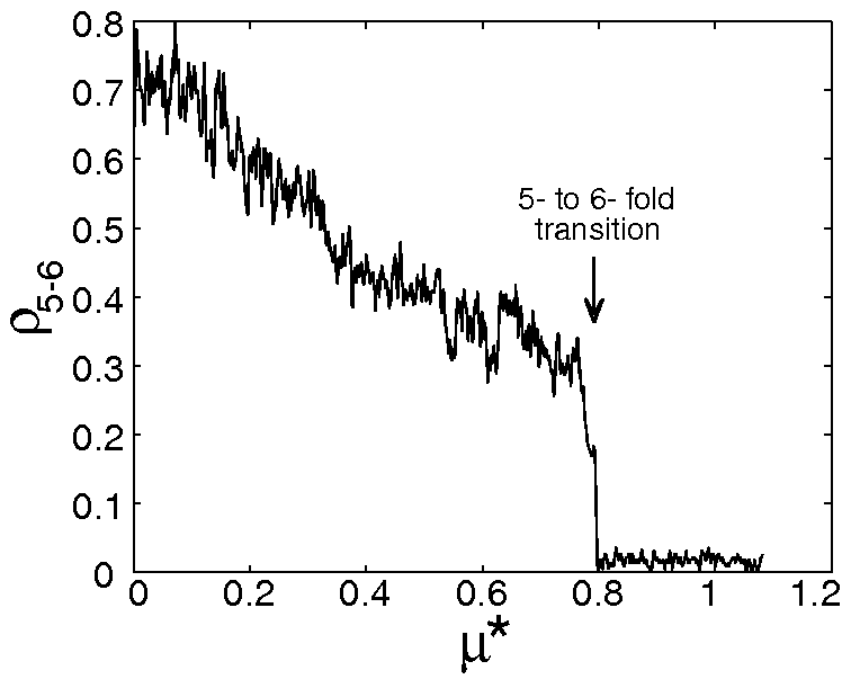

**Figure 3.** Order parameter $r_{5-6}$ vs. reduced chemical potential $\mu^*$ for Xe on Al-Ni-Co along the monolayer plateau of the 77 K isotherm. The commensurate-incommensurate transition from the five-fold to the six-fold structure is evident as a step at a µ• value of about 0.8.

Interestingly, similar graphs for the other rare gases indicate that *no* transition occurs[11], even after $2^{nd}$ and $3^{rd}$ layer adsorption. That is, the film never converts completely to a hexagonal structure in these cases. On a completely flat surface, lacking atomicity, the 2D solid structure is expected to be hexagonal at all coverages. Therefore, while the corrugation of the quasicrystal delays the hexagonal ordering of Xe until nearly a monolayer is present, it precludes the hexagonal ordering for the smaller rare gases for all coverages up to several layers.

The conditions under which an incommensurate structure forms is particularly relevant to interfacial friction [14]. The results of the simulations described here indicate that the size of the gas atoms is the most important factor for determining whether an incommensurate structure will form. More precisely, the mismatch in size between the adsorbate and the substrate determines whether a transition will occur. The potential shown in Figure 1 shows an aperiodic corrugated surface. Although the surface is aperiodic, there are paths of relatively low energy (troughs). Figure 1 shows 3 paths through this potential. We expect that the gas atoms preferentially choose these troughs for adsorption sites. While the troughs themselves are aperiodically spaced, the average spacing between them is 0.38 nm, which is exactly the inter-row distance present in bulk Xe. Therefore, if Xe atoms line up in the troughs, their inter-row distance is exactly that of bulk Xe planes, and therefore the substrate facilitates the formation of a hexagonal Xe layer. This is not the case for the smaller rare gases.

**Table 3.** Adsorption potential parameters for a variety of Xe-like gases. SD is the standard deviation of the variation of adsorption energy, a measure of the magnitude of the corrugation.

| Gas | $\sigma_{gs}$ (nm) | SD (meV) | $\sigma_{gs}$/ SD | transition |
|---|---|---|---|---|
| dXe(1) | 0.260 | 29.18 | 112 | no |
| dXe(2) | 0.316 | 19.33 | 61 | no |
| Xe | 0.326 | 17.93 | 55 | yes |
| iXe(1) | 0.396 | 11.21 | 28 | yes |
| iXe(2) | 0.458 | 7.77 | 17 | yes |

This fortuitous conjunction of natural spacings for Xe and this Al-Ni-Co surface is clearly an important condition for the formation of the hexagonal layer for Xe. The corrugation of the Al-Ni-Co potential prevents the transition for smaller gases. But does the hexagonal ordering rely on matching the gas spacing to the QC spacing? In order to address this, simulations were performed on fictitious gases that have the same average holding potential and same gas-gas interactions as Xe, but different sizes. The parameters for these gases are shown in Table 3. It was found that for gases that are at least as large as Xe, there is a transition into the hexagonal phase, indicating that the transition does not rely solely on the size match. For these gases, the effective corrugation of the potential is apparently too small to prevent the natural formation of a hexagonal layer.

In summary, the studies described here show that for rare gas adsorption
(1) The Al-Ni-Co potential surface experienced by the adsorbate is highly corrugated compared to that on typical metal surfaces.
(2) The average heights of the gas atoms above the Al-Ni-Co surface are smaller than the heights expects for the gas atoms above flat surfaces having the same average holding potential.
(3) The corrugation precludes the hexagonal ordering of rare gases than are smaller than Xe.

This last point is partially due to the coincidental size match between Xe and this QC surface, but gases larger than Xe also form hexagonal structures at sufficiently high chemical potentials.

We gratefully acknowledge useful conversations with Chris Henley. This research was supported by NSF grants DMR-0208520 and DMR-0505160. We also acknowledge the San Diego Supercomputer Center for computing time under Grant Number MSS060002.

# References

[1] R. McGrath, J. Ledieu, E. J. Cox, and R. D. Diehl, J. Phys.: Condensed Matter **14** R119 (2002).

[2] V. Fournée and P. A. Thiel, J. Phys. D: Appl. Phys. **38** R83 (2005).

[3] N. Ferralis, R. D. Diehl, K. Pussi, M. Lindroos, I. R. Fisher, and C. J. Jenks, Phys. Rev. B **69** 075410 (2004).

[4] S. Curtarolo, W. Setyawan, N. Ferralis, R. D. Diehl, and M. W. Cole, Phys. Rev. Lett. **95** 136104 (2005).

[5] R. D. Diehl, N. Ferralis, K. Pussi, M. W. Cole, W. Setyawan, and S. Curtarolo, Philos. Mag. **86** 863 (2006).

[6] W. Setyawan, N. Ferralis, R. D. Diehl, M. W. Cole, and S. Curtarolo, Phys. Rev. B **74** (2006).

[7] R. D. Diehl, J. Ledieu, N. Ferralis, A. W. Szmodis, and R. McGrath, Journal of Physics: Condensed Matter **15** R63 (2003).

[8] P. Zeppenfeld, in *Landolt Boernstein: Physics of Covered Surfaces*, Vol. III/42A (H. P. Bonzel, ed.), Springer, Berlin, 2001, p. 67.

[9] R. A. Trasca, N. Ferralis, R. D. Diehl, and M. W. Cole, J. Phys: Condens. Mat. **16** S2911 (2004).

[10] N. Ferralis, K. Pussi, E. J. Cox, M. Gierer, J. Ledieu, I. R. Fisher, C. J. Jenks, M. Lindroos, R. McGrath, and R. D. Diehl, Phys. Rev. B **69** 153404 (2004).

[11] W. Setyawan, N. Ferralis, R. D. Diehl, M. W. Cole, and S. Curtarolo, J. Phys.: Condens. Matt. to be published (2006).

[12] L. E. Cascarini de Torre and E. J. Bottani, Langmuir **13** 3499 (1997).

[13] J. L. F. Da Silva, C. Stampfl, and M. Scheffler, Phys. Rev. B **72** 075424 (2005).

[14] B. N. J. Persson and E. Tosatti, eds., *The Physics of Sliding Friction*, Kluwer Academic, Dordrecht, 1996.

[15] R. D. Diehl, T. Seyller, M. Caragiu, G. S. Leatherman, N. Ferralis, K. Pussi, P. Kaukasoina, and M. Lindroos, J. Phys.: Condens. Matt. **16** S2839 (2004).